\documentclass[reprint,amsmath,amssymb,aps,prl,floatfix,longbibliography]{revtex4-2}
\usepackage{graphicx}
\usepackage[bookmarks=false,hidelinks]{hyperref}

\begin{document}
\title{Observation of Universal Quench Dynamics and Townes Soliton Formation from
Modulational Instability in Two-Dimensional Bose Gases}
\author{Cheng-An Chen$^{ 1 }$ and Chen-Lung Hung$^{1, 2 ,}$}
\email{clhung@purdue.edu}
\address{$^1$ Department of Physics and Astronomy, Purdue University, West Lafayette, Indiana 47907, USA}
\address{$^2$ Purdue Quantum Science and Engineering Institute, Purdue University, West Lafayette, Indiana 47907, USA}

\begin{abstract}
We experimentally study universal nonequilibrium dynamics of two-dimensional atomic Bose gases quenched from repulsive to attractive interactions. We observe the manifestation of modulational instability that, instead of causing collapse, fragments a large two-dimensional superfluid into multiple wave packets universally around a threshold atom number necessary for the formation of Townes solitons. We confirm that the density distributions of quench-induced solitary waves are in excellent agreement with the stationary Townes profiles. Furthermore, our density measurements in the space and time domain reveal detailed information about this dynamical process, from the hyperbolic growth of density waves, the formation of solitons, to the subsequent collision and collapse dynamics, demonstrating multiple universal behaviors in an attractive many-body system in association with the formation of a quasistationary state.
\end{abstract}
\maketitle

Predicting the evolution of multidimensional nonlinear systems under attractive interactions is a challenging task, owing to the instability to collapse \cite{Robinson1997,donley2001dynamics, moll2003self}.
Bright solitons are remarkable stationary states, established when the self-focusing effect responsible for collapse is exactly compensated by wave dispersion \cite{zabusky1965interaction,shabat1972exact}. In uniform two-dimensional (2D) systems with standard cubic interactions, such as Kerr medium \cite{chiao1964self,moll2003self} or matter waves formed by weakly interacting 2D Bose gases \cite{chomaz2015emergence,Hung2011}, however, such intricate balance cannot be fulfilled except when a wave packet possesses a critical norm (or total atom number) known as the Townes threshold and a specific waveform known as the Townes profile \cite{chiao1964self,kartashov2011solitons,malomed2016multidimensional} -- only at which bright solitons can form. A Townes soliton is predicted to be unstable as long as the norm deviates from the Townes threshold \cite{moll2003self,kartashov2011solitons}. Despite extensive interest in multidimensional bright solitons \cite{fleischer2003observation,Cornish2006_3Dsoliton,kartashov2011solitons,malomed2016multidimensional,kartashov2019frontiers}, including recent advancements on 2D spatial solitons in various nonlinear optics settings \cite{fleischer2003observation,moll2003self,chen2012optical}, an experimental realization and characterization of Townes solitons has remained elusive.

In soliton formation dynamics, modulational instability (MI) is a ubiquitous mechanism that causes the amplification of initial wave disturbances and fragmentation into solitary waves \cite{ Robinson1997,zakharov2009modulation,nguyen2017formation,everitt2017observation}. In one-dimensional (1D) systems, MI is responsible for the formation of stable soliton trains, for example, in nonlinear fiber optics \cite{tai1986observation,solli2012fluctuations,dudley2014instabilities}, in 1D Bose gases \cite{strecker2002formation,khaykovich2002formation,nguyen2017formation,everitt2017observation}, or in Bose-Einstein condensates in optical lattices \cite{KonotopMI2002,carusotto2002nonlinear,Morsch2006}. In higher spatial dimensions, transverse MI and wave fragmentation were studied in various types of bulk nonlinear optical media \cite{Mamaev1996,kivshar2000self,chen2012optical}. However, detailed dynamics of multidimensional MI and its possible connection to the universal formation of a quasistationary state, the unstable Townes solitons, have not been clearly demonstrated.

\begin{figure}[t]
\centering
\includegraphics[width=1\columnwidth]{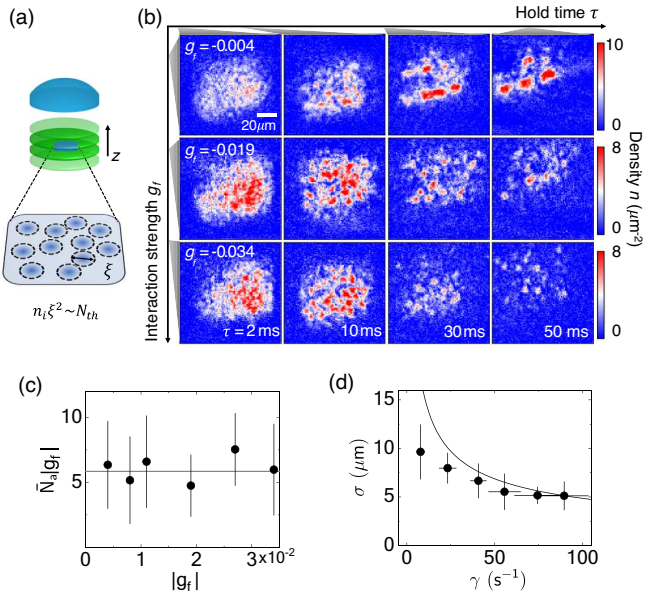}
\caption{Universal solitary wave formation in quenched 2D Bose gases. (a) Interaction quench-induced MI fragments a 2D gas (blue shaded square) into wave packets of a characteristic size $\xi$ (dashed circles) that contains atom number $\sim n_i\xi^2$ approaching the Townes threshold $N_\mathrm{th}$. The 2D gas is confined in a single node of a repulsive standing-wave potential (green shaded ovals), evolves for a hold time $\tau$, and is imaged via a microscope objective (blue hemisphere) \cite{SM}. (b) Single-shot images of samples quenched to the indicated interaction $g_f$ (in each row) and held for the labeled time $\tau$ (in each column). Solitary waves (isolated density blobs) become visible at $\tau \geq 30~$ms. (c) Scaled mean atom number in a solitary wave $\bar{N}_a|g_f|$. Solid line marks the universal threshold $N_\mathrm{th}|g_f| = 5.85$. (d) Mean size $\sigma$ versus interaction frequency $\gamma$. Solid line is the interaction length $\xi$. All data points in (c) and (d) are evaluated at $\tau = 42\sim 50~$ms except for those of $g_f=-0.004$ which are evaluated at $\tau=150\sim 200$~ms. Error bars are standard errors. Uncertainties in $g_f$ are smaller than the size of symbols \cite{SM}.}
\label{fig:demo}
\end{figure}

Using ultracold 2D Bose gases, we show that universal MI dynamics supports the critical formation of Townes solitons. Starting with a 2D superfluid of an initial density $n_i$ and quenched to an attractive interaction $g_f<0$, we show that MI causes collective modes with a wave number around $k_\mathrm{MI}=\pi/\xi$ associated with the interaction length $\xi=\pi/\sqrt{2n_i|g_f|}$ to grow predominantly \cite{Salasnich2003MI,Carr2004MI,guth2015dark}, fragmenting the superfluid as illustrated in Fig.~\ref{fig:demo}(a). Intriguingly, the characteristic atom number in each wave packet $\sim n_i\xi^2=\pi^2/2|g_f|$ well approaches the Townes threshold $N_\mathrm{th} = 5.85/|g_f|$ \cite{malomed2016multidimensional}, thus opening up a pathway for Townes soliton formation. This relation should apply universally for any $n_i$ and $g_f$ provided no other length scales set in. We note that, due to the scaling symmetry in 2D, a Townes soliton forms under a scale-invariant profile \cite{chiao1964self} and MI can set the physical length scale $\xi$ that depends only on the product of $n_i$ and $|g_f|$.

In this Letter, we report the observation of universal dynamics and Townes soliton formation in quenched 2D Bose gases. We observe MI that breaks up an otherwise large 2D sample into fragments universally around the Townes threshold. We clearly identify solitary waves whose density distributions agree well with the Townes profiles -- the stationary state solution of the 2D nonlinear Schr\"{o}dinger equation (NLSE) \cite{chiao1964self,malomed2016multidimensional}. Our measurement further reveals universal solitary wave dynamics governed by the MI timescale and a universal scaling behavior in the density power spectra, allowing us to clearly identify a distinct time period for the unstable growth of density waves (while conserving total atom number), followed by a short era of wave collapse and soliton formation.

We begin the experiment with a uniform 2D Bose gas deep in the superfluid regime \cite{SM}, which is formed by $N\approx 1.5\times 10^4$ cesium atoms trapped inside a quasi-2D box potential. The atomic surface density is $n_i \approx 5/\mu$m$^2$ and the surface area is controlled by a wall-like potential that is removed following the interaction quench \cite{SM}. The tight vertical ($z$) confinement of the box freezes all atoms in the harmonic ground state along $z$ axis. The trap vibrational level spacing ($\omega_z = 2\pi \times 1750~$Hz) is more than 2 orders of magnitude larger than the attractive interaction energy studied, ensuring that the observed wave dynamics is effectively 2D \cite{pedri2005two}. The interaction strength $g =\sqrt{8\pi}a/l_z$ is controlled by the tunable s-wave scattering length $a$ via a magnetic Feshbach resonance \cite{chin2010feshbach}, and $l_z=208~$nm is the vertical harmonic oscillator length; $g=g_i=0.115$ is the initial interaction strength.

We perform an interaction quench (in 1~ms) to various $g=g_f$ in samples with a large surface area $A\approx (60~\mu$m$)^2$. Following a variable hold time $\tau$, we perform single-shot absorption imaging to record the sample density distribution as shown in Fig.~\ref{fig:demo}(b). Around 30 samples are imaged for ensemble analyses. In a short hold time, we observe density blobs randomly clumping up throughout a sample. The sizes of the blobs are smaller with larger $|g_f|$. At a longer hold time, $\tau \geq 30~$ms, the number of observed blobs reduces, becoming more isolated, although the mean size and atom number of the surviving blobs remain nearly unchanged for $\tau \leq 200~$ms (see Fig.~S2 in Ref. \cite{SM}). The same quench protocol applied to samples in a three-dimensional trap ($\omega_z \approx 2\pi\times 100~$Hz with a weak radial trap frequency $\omega_r \approx 2\pi\times13$~Hz) results in no observed density blobs.

We characterize these isolated blobs (solitary waves) and compare their atom number with the Townes threshold. We approximate their density profiles by 2D Gaussians \cite{malomed2016multidimensional} and fit the mean size $\sigma$ and atom number $\bar{N}_a$ \cite{SM}. Within the interaction range studied, $-0.004\geq g_f\geq -0.034$, in Fig.~\ref{fig:demo}(c) we find that all rescaled atom numbers $\bar{N}_a|g_f|$ fall around $N_\mathrm{th}|g_f|=5.85$, giving a mean $\overline{\bar{N}_a|g_f|}=6(1)$. Interestingly, the standard deviation of the atom number $\delta N_a$ scales with $g_f$ accordingly, giving a mean fluctuation $\overline{\delta N_a|g_f|}=$3.2(5)$\sim 0.5\overline{\bar{N}_a|g_f|}$. We believe that the number variation around the threshold results from the energy-time uncertainty relation, as we later show that these blobs form in a timescale $\sim \gamma^{-1}$, where $\hbar\gamma=\hbar^2n_i|g_f|/m$ is the interaction energy, $m$ is the atomic mass, and $\hbar$ is the reduced Planck constant. Moreover, the size of these solitary waves also agrees with the prediction $\sigma \approx \xi$ in Fig.~\ref{fig:demo}(d), indicating that MI provides the length scale for the formation of blobs. A size discrepancy at $\gamma \approx 2\pi\times 1.2$~Hz ($g_f=-0.004$) could likely be attributed to the influence of a very weak horizontal corrugation in the vertical confining potential.

\begin{figure}[t]
\centering
\includegraphics[width=0.95\columnwidth]{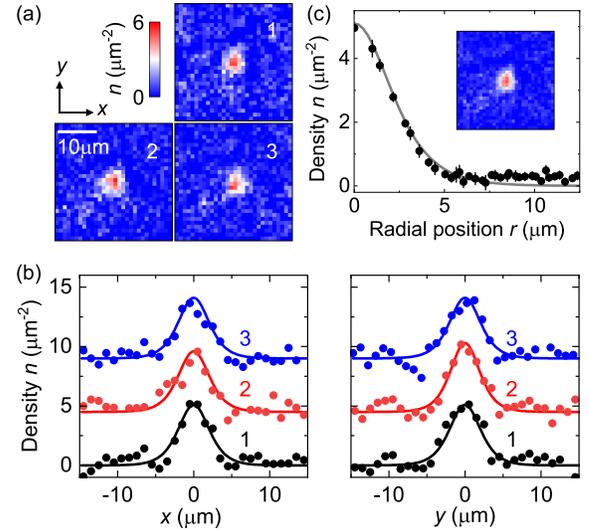}
\caption{Townes solitons and the Townes profiles. (a) Sample images of single solitary waves ($g_f=-0.034$) recorded at $\tau=100~$ms. (b) Density line cuts (solid circles) through the center of images as numerically labeled in (a). Each data is offset by $4.5/\mu$m$^2$ for viewing. Solid lines are the Townes profiles of peak densities $n_0=5.1/\mu$m$^2$ (for \#1,\#3) and 5.8$/\mu$m$^2$ (for \#2), respectively. (c) Azimuthally averaged radial profile (solid circles) from the mean density image of (a) (inset: $30\times 30~\mu$m$^2$), in close comparison with theory (solid curve) calculated using $n_0=5.1/\mu$m$^2$. Nearby dispersed blobs contribute to a small background at $r\gtrsim 7\mu$m.} \label{fig:profile}
\end{figure}

To confirm that the quench-induced solitary waves indeed form Townes solitons, we compare their density distributions with the scale-invariant, isotropic Townes profile $n(r)=\alpha^2|\phi (\alpha r)|^2/(2|g_f|)$, where $\alpha = \sqrt{2n_0 |g_f|/|\phi (0)|^2}$ is a scale factor. Given the peak density $n_0$, the characteristic size and density profile of a Townes soliton are uniquely determined. Here, the radial function $\phi(\tilde{r})$ is the stationary solution of the scale-invariant 2D NLSE \cite{chiao1964self,malomed2016multidimensional}. The stationary profile $\phi(\tilde{r})$ has been evaluated numerically \cite{chiao1964self}, giving $|\phi(0)|\approx 2.207$, and the norm $\int |\phi(\tilde{r})|^2d\tilde{\mathbf{r}} \approx 11.7$ sets the Townes threshold.

Since MI sets the length scale during the soliton formation process, $\alpha \sim \xi^{-1}$, the peak density of most solitons should be comparable to the initial sample density $n_0\approx n_i$. Figure~\ref{fig:profile} plots three isolated solitary waves of similar peak density that are randomly chosen from quenched samples ($g_f=-0.034$) at a long hold time $\tau=100~$ms. Their individual density distributions, as well as the averaged radial density profile, indeed agree fairly well with the expected Townes profiles with no fitting parameters. More agreement with the Townes profiles is discussed in Ref. \cite{SM}, where a single array of well-isolated solitons can form in an elongated sample following an interaction quench. Our observation confirms that Townes solitons can prevail from MI and explains why many randomly formed solitary waves, as observed in Fig.~\ref{fig:demo}, are long-lived. We note that, in the absence of MI and fragmentation, a wave was observed to collapse only partially to a Townes profile~\cite{moll2003self}.

\begin{figure}[t]
\centering
\includegraphics[width=1\columnwidth]{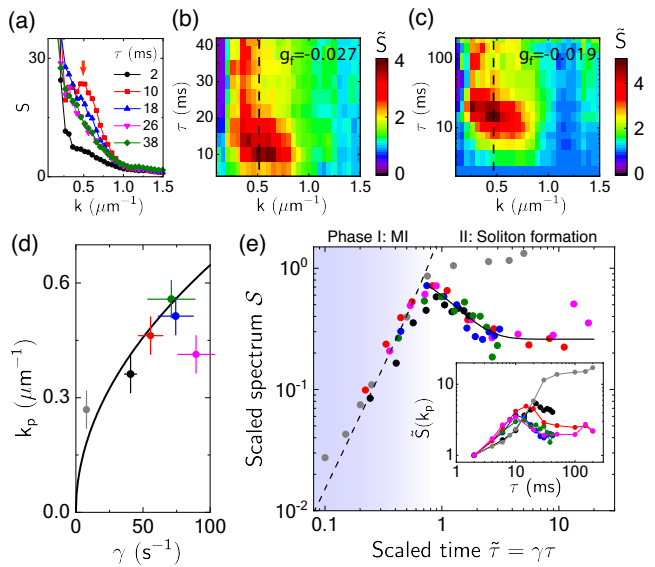}
\caption{Dynamics and universal scaling in the density power spectra. (a) Sample spectra $S(k,\tau)$ at $g_f=-0.027$ and the indicated hold time $\tau$. (b) Corresponding growth spectra $\tilde{S}(k,\tau)=S(k,\tau)/S(k,\tau_0)$ with $\tau_0=2~$ms. (c) Sample growth spectra $\tilde{S}$ at $g_f=-0.019$ and hold time up to 200~ms. Vertical dashed lines in (b) and (c) mark the peak wave number $k_p$. (d) $k_p$ versus interaction frequency $\gamma$ (filled circles) measured at $g_f= -0.004$ (gray), $-0.011$ (black), $-0.019$ (red), $-0.027$ (blue), and $-0.034$ (magenta and olive), respectively. Corresponding $\tilde{S}(k_p,\tau)$ are shown in the inset of (e). Solid line is the prediction $k_\mathrm{MI}=\sqrt{2\gamma m/\hbar}$. Error bars include systematic and statistical errors. (e) Scaled spectra $\mathcal{S}$ plotted using Eq.~(\ref{eq:scale}), which collapse approximately onto a single curve except for the one at $g_f=-0.004$. Dashed line (Solid line) is a hyperbolic (exponential) fit to Phase I, $\tilde{\tau}<0.8$ (II, $\tilde{\tau}>0.8$), of the collapsed spectra; see text.}
\label{fig:spectrum}
\end{figure}

We now turn to study the universal dynamics during the soliton formation process. We focus on the time evolution of density power spectrum \cite{Hung2013} $S(k,\tau) = \langle |n(k,\tau)|^2\rangle/N$ in the spatial frequency domain (momentum space), using the Fourier transform of the sample density distribution. Here $k=|\mathbf{k}|$ is the momentum wave number of the azimuthally averaged spectrum, $N$ is the total atom number, and $\langle \cdots \rangle$ denotes ensemble averaging. In the power spectra [Fig.~\ref{fig:spectrum}(a)], we clearly observe rapid growth of a nonzero momentum peak at short hold time (marked by an arrow), indicating the development of density waves at a dominant length scale throughout the sample. The nonzero momentum peak then dissipates at longer hold time until the power spectrum finally becomes monotonic and stationary, which signifies the collapse of density waves and fragmentation of the sample into solitons that later becomes uncorrelated in coordinate space.

The evidence of MI-induced wave amplification at different interaction strengths is best illustrated when we plot the relative growth spectra $\tilde{S}(k,\tau) = S(k,\tau)/S(k,\tau_0)$, normalized by the initial power spectrum at $\tau_0=2~$ms. This allows us to determine which mode has the largest growth rate. For different samples in Figs.~\ref{fig:spectrum}(b) and \ref{fig:spectrum}(c), the momentum peak is clearly visible within $0.2/\mu$m$<k<1/\mu$m at short hold time. The growth patterns look similar for samples with different $g_f$, although the peak location, height, and the evolution timescale vary. We identify the peak wave number $k_p$ and find consistency with the prediction from MI in Fig.~\ref{fig:spectrum}(d).

Another remarkable prediction from MI is that, regardless of the dimensionality of the system, the power spectrum at $k_p$ exhibits a universal time and amplitude scaling behavior with respect to the interaction timescale $\gamma^{-1}$. This is based on extending the Bogoliubov phonon picture to the regime under attractive interactions, which predicts that collective modes of opposite momenta are generated in pairs, as seeded from initial density perturbations, and subsequently form density waves along the associative directions while being amplified at a rate $\gamma$ until significant depletion of the ground state atoms occurs \cite{SM}.

In Fig.~\ref{fig:spectrum}(e), we experimentally test the scaling relation in the peak growth spectra $\tilde{S}(k_p,\tau)$, covering the entire time period. We summarize the scaling relation as follows:
\begin{equation}
    \mathcal{S}(\tilde{\tau}) \approx \frac{\gamma}{\bar{\gamma}_i} [\tilde{S}(k_p,\tilde{\tau})-1],\label{eq:scale}
\end{equation}
where $\tilde{\tau}=\gamma\tau$ is the scaled time and $\mathcal{S}(\tilde{\tau})$ is the scaled spectrum. In the amplitude scaling, we normalize $\gamma$ with the mean initial interaction energy unit $\bar{\gamma}_i=306~$s$^{-1}$ before the quench \cite{SM}. We show that six power spectra, each with different $\gamma$, can collapse onto a single curve over a surprisingly long scaled time $\tilde{\tau}<10$. The only exception is the spectrum at $g_f=-0.004$, where we have used $\gamma^*= 3.2\gamma$ to force its collapse within $\tilde{\tau}\leq 0.8$. This different behavior is likely due to a weak horizontal trap corrugation influencing the dynamics, as noted in Fig.~\ref{fig:demo}(d).

From this universal spectrum, we identify two distinct regimes of dynamics, divided by a critical time $\tilde{\tau}_c \approx 0.8$ as shown in Fig.~\ref{fig:spectrum}(e). We label the time period $ \tilde{\tau} \leq \tilde{\tau}_c$ for MI with an amplified (hyperbolic) growth of density waves \cite{SM},
\begin{equation}
\tilde{S}(k_p,\tau) \approx 1 + \alpha \frac{\bar{\gamma}_i}{\gamma} \sinh^2(\gamma \tau),
\end{equation}
where $\alpha=1.5(1)$ is determined from a fit to $\mathcal{S}(\tilde{\tau})$ for $\tilde{\tau}\leq 0.5$; $\alpha=2$ is obtained from the theory calculation for $\tilde{\tau}\ll1$, neglecting the depletion of ground state atoms, dissipation, or interaction between the collective modes. Beyond $ \tilde{\tau} \geq \tilde{\tau}_c$ after $\mathcal{S}(\tilde{\tau})$ reaches the order of unity, dynamics enters the second phase, decaying with a time constant $\Delta\tilde{\tau}\sim 0.8$ and transitioning to a slowly evolving, quasistationary behavior.

Our data suggest the existence of a universal time and amplitude scaling behavior and a limit for the amplified density wave, followed by a universal dynamics for the wave collapse and soliton formation. For $g_f=-0.004$, however, $\tilde{S}(k_p,\tau)$ remains slowly growing within $1\leq \gamma^*\tau  \leq 10$, suggesting a less severe wave collapse.

\begin{figure}[t]
\centering
\includegraphics[width=0.95\columnwidth]{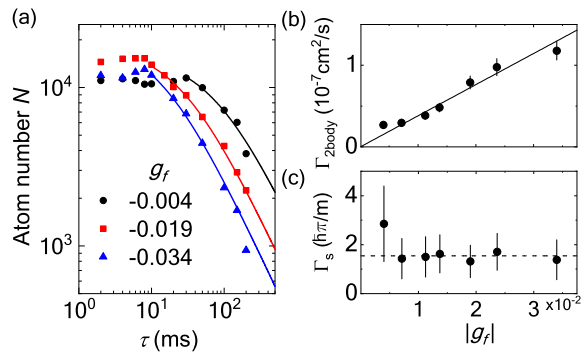}
\caption{Universal soliton collision dynamics in 2D. (a) Sample loss curves in total atom number $N$ measured at the indicated interaction $g_f$. Solid lines are two-body loss fits after atom loss initiates. (b) Fitted rate coefficients. Solid line is a linear fit, giving a slope $\Gamma_\mathrm{2body}/|g_f|=3.8(2)\times 10^{-6}$cm$^2/$s. (c) 2D soliton binary loss coefficients $\Gamma_s$ determined from the rate coefficients in (b) and $\bar{N}_a$ as in Fig.~\ref{fig:demo}(c), and compared with the universal prediction Eq.~(\ref{eq:collision}), giving a mean $\eta=1.5(1)$ (dashed line) and agreeing with $\eta=1.5(3)$ alternatively determined from the fitted slope in (b) and mean $\overline{\bar{N}_a|g_f|}=6(1)$. Error bars are standard errors. Uncertainties in $g_f$ are smaller than the size of symbols \cite{SM}.}
\label{fig:loss}
\end{figure}

Following the observed density wave collapse dynamics, a 2D sample fragments into many solitary wave packets of characteristic size $\sigma\sim\xi$. As seen in Fig.~\ref{fig:demo}, due to the close proximity of many wave packets (also with characteristic distances $\sim \xi$), collisions between them may induce collapses that lead to rapid atom number loss. Here, we show that the MI timescale continues to dominate the collision dynamics and the total atom number loss in a quenched 2D Bose gas.

In a recent study of 1D soliton collisions \cite{nguyen2014collisions}, it is shown that merger occurs when solitons of similar phases collide, and two solitons of opposite phases appear to repel each other. In 2D, the merger of two solitonlike wave packets should lead to a new atom number $N_a >N_\mathrm{th}$. This can induce collapse that quickly removes the merged soliton. For solitons or density blobs formed by MI with randomly seeded density waves in a large 2D sample, one expects no fixed phase relationship between neighbors. Merger can thus occur randomly throughout the sample and the total atom number loss may reveal the scaling of soliton binary collision losses.

In Fig.~\ref{fig:loss}, we examine the total atom number loss for the samples quenched to different $g_f$ (Fig.~\ref{fig:demo}). We observe the onset of loss in each sample at a time $\tau$ corresponding roughly to the critical time $\tilde{\tau}_c$, a behavior similarly observed for MI in 1D \cite{nguyen2017formation}. Beyond the critical time, we confirm that the loss curves can be well-captured by a simple two-body loss model, $\dot{N}/N = -\Gamma_\mathrm{2body} N/A$. We attribute this behavior to the dominance of binary collisions between solitons or density blobs that trigger collapse and atom number loss; without triggered collapse, the usual three-body recombination loss should be negligible \cite{kraemer2006evidence}. In Fig.~\ref{fig:loss}(b), we find a linear dependence on $|g_f|$ in the measured loss coefficients $\Gamma_\mathrm{2body}$. This in fact suggests a constant binary loss coefficient $\Gamma_s$ for colliding wave packets, where $\Gamma_s=\Gamma_\mathrm{2body}\bar{N}_a$ \cite{SM} and $\bar{N}_a|g_f|\approx6$ is the measured universal number for solitary waves formed by MI.

We suspect this universal loss behavior results from MI scaling and 2D scale invariance, which suggests a constant binary loss coefficient,
\begin{equation}
    \Gamma_s = \eta \frac{\hbar\pi}{m},\label{eq:collision}
\end{equation}
independent of the interaction parameters ($n_i, g_f)$. This is because the collision rate $\Gamma_s\sim \sigma\bar{v}$ and the dependences on length scales in the linear cross section $\sigma \approx \xi$ and relative velocity $\bar{v}\approx \sqrt{2}\hbar\pi/m\xi$ cancel each other; the constant $\eta\approx \sqrt{2}$ is estimated for $\sim$50\% probability of merger per collision or, equivalently, on average one soliton or density-blob loss per collision event. If Eq.~(\ref{eq:collision}) holds, we expect a collision lifetime $1/n_s\Gamma_s\sim \gamma^{-1}$, where $n_s=n_i/\bar{N}_a$ is the initial soliton/blob density \cite{SM}. This suggests that wave collapse and binary collision likely take place at the same rate during the second phase of the density evolution.

To unambiguously confirm the universality of collision dynamics, in Fig.~\ref{fig:loss}(c) we deduce $\Gamma_s$ independently using experimentally determined values ($\Gamma_\mathrm{2body}$, $\bar{N}_a$) at each $g_f$. Our results conform very well with the prediction by Eq.~(\ref{eq:collision}), giving a mean $\eta \approx 1.5$. We emphasize here that the loss coefficients universally depend only on the physical constants $\hbar/m$ is a remarkable manifestation of MI and scale-invariant symmetry in 2D. The observations in Figs.~\ref{fig:spectrum} and \ref{fig:loss} together confirm that interaction quench dynamics leads to Townes soliton formation at $\tau\gtrsim \gamma^{-1}$, followed by collision (that induces collapse) also at the same timescale $\gamma^{-1}$ universally governed by MI.

In summary, we study the universal nonequilibrium dynamics in degenerate 2D Bose gases quenched from repulsive to attractive interactions, and observe the dynamical formation of Townes solitons from MI. Townes solitons are observed to be collisionally unstable. However, further stabilization and manipulation may be possible \cite{Sait2003,Efremidis2003,Baizakov2004_periodic, fleischer2003observation}. We note that the initial shape and finite sample size can be further manipulated to invoke strong boundary effects in quench-induced MI \cite{SM}. Using slow interaction ramps may also induce MI dynamics deviating from the reported universal behavior. Soliton formation may be disrupted, leading to, for example, only partial collapse to the Townes profiles \cite{moll2003self}. Lastly, we comment that controlled formation of 2D solitons via pair-production processes in MI may find future applications in matter-wave interferometry \cite{cronin2009optics,lucke2011twin}, or even in the generation and distribution of many-body entanglement \cite{lange2018entanglement,kunkel2018spatially,fadel2018spatial,shin2019bell}.

We acknowledge discussions with Qi Zhou, Chih-Chun Chien, Sergei Khlebnikov, and Chris Greene. We thank Cheng Chin for discussions and many critical comments. We are grateful to H. J. Kimble for instrument loan since the early stage of this project. We thank May Kim, Yiyang Fang, and Wuxiucheng Wang for laboratory assistance. This project is supported in part by the Purdue Research Foundation, the W. M. Keck Foundation, the NSF Grant No. PHY-1848316, and the U.S. Department of Energy (Grant No. DE-SC0019202).

\appendix
\renewcommand{\figurename}{Fig.}
\renewcommand{\thefigure}{S\arabic{figure}}
\setcounter{figure}{0}
\renewcommand{\theequation}{S\arabic{equation}}
\setcounter{equation}{0}
\setcounter{page}{1}

\section{Supplemental Material}
\subsection{Experimental procedures}\label{SM:Exp}
\subsubsection{Formation of uniform 2D Bose gases}
\begin{figure}[t]
\centering
\includegraphics[width=1\columnwidth]{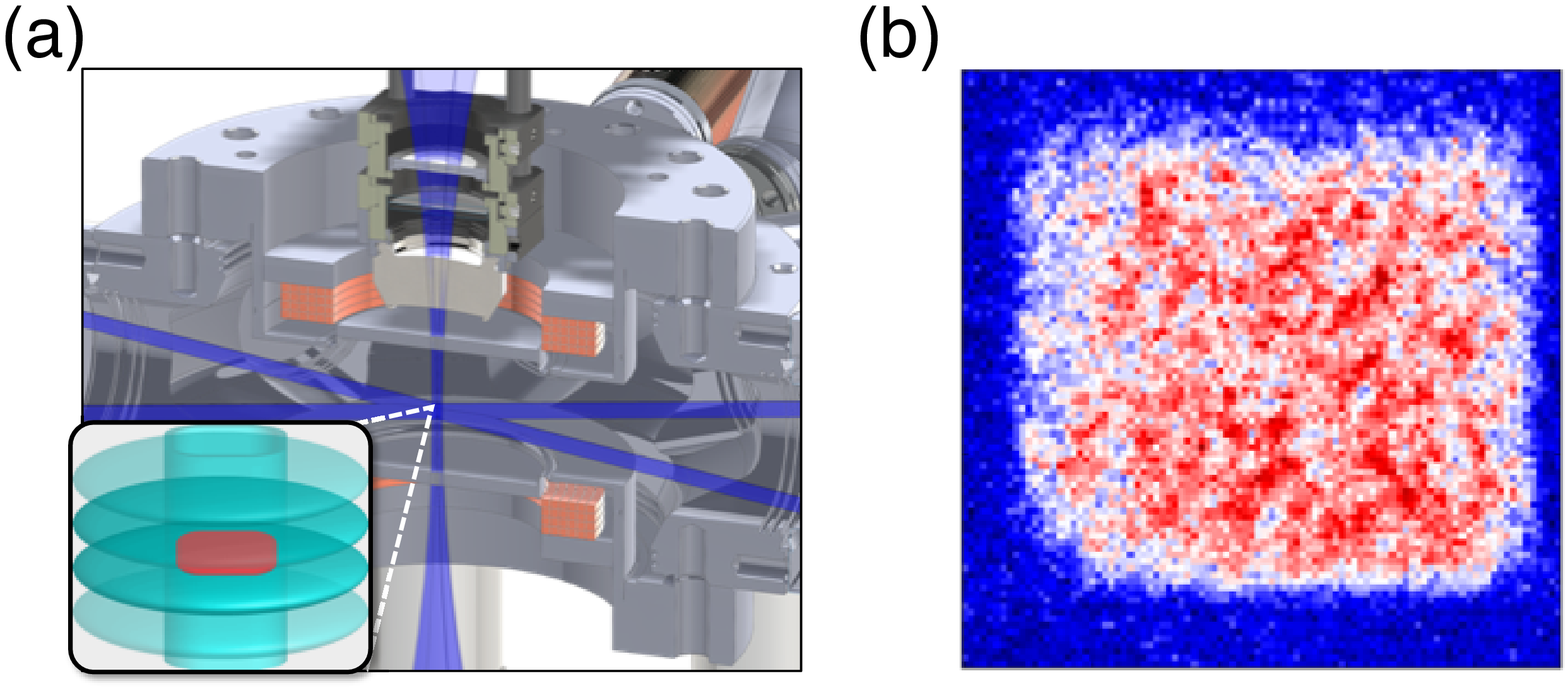}
\caption{Schematic of the experiment. (a) Two crossed vertical confinement beams (blue shaded sheets) and the wall beam (blue shaded cylinder). Inset illustrates a 2D gas (red shaded square) confined in the box potential, which is represented by a single node between the blue shaded ovals with the square-wall cylinder beam forming the box boundary. (b) Single-shot sample image of a 2D gas loaded in the box potential. Image size is 90$\times 90~\mu$m$^2$ and pixel size is $1~\mu$m$^2$.}
\label{figSM:scheme}
\end{figure}

Figure~\ref{figSM:scheme} shows the schematics of the experiment. Our uniform 2D Bose gas is confined in a quasi-2D box potential formed by all repulsive optical dipole beams. The vertical box confinement is provided by a single node of a repulsive standing-wave potential (2$~\mu$m periodicity) along the vertical ($z$-) direction. The node defines the horizontal 2D plane. The measured vertical trap frequency in the node is $\omega_z/2\pi =1.75(2)~$kHz $\gg (k_B T, |\mu|)/\hbar$, freezing all atoms in the harmonic ground state. Here $k_B$ is the Boltzmann constant, $T\leq8~$nK is the temperature measured in time-of-flight, and $\mu$ is the chemical potential. The horizontal boundary of the box is formed by a tightly focused repulsive optical beam that scans the box boundary to form a time-averaged wall potential. The beam is controlled by a pair of acousto-optic deflector and is projected through the same microscope objective (numerical aperture N.A. = 0.37) used for imaging. Optical resolution of our imaging system is $\lesssim 2\mu$m, determined using the smallest density blobs identified in the experiment.

The 2D Bose gas is prepared deeply in the 2D superfluid regime with a phase space density $n_i\lambda_\mathrm{dB}^2 \gtrsim 15$ \cite{Hung2011}. Here $n_i\approx 5/\mu$m$^2$ is the initial 2D density and $\lambda_\mathrm{dB}$ is thermal de Broglie wavelength. Finite temperature mainly contributes initial thermal phonon fluctuations in a superfluid sample that seed the growth of density waves from modulational instability.

We perform the interaction quench while simultaneously removing the wall potential to allow hot collision products (due to soliton collapses) to eject out of the experiment field of view. When the wall potential is removed, the residual global trap frequency in the horizontal plane, due to a weak magnetic trap, is experimentally determined to be $\omega_r /2\pi < 1.5~$Hz through superfluid dipole oscillations.

\subsubsection{Magnetic two-body interaction tuning}
The scattering length is tuned using a magnetic Feshbach resonance \cite{chin2010feshbach,kraemer2006evidence}. We identify zero scattering length at the magnetic field $B=17.120(6)$G, using the superfluid in-situ size as well as the expansion rate in a 2D time-of-flight. We adopt the formula \cite{kraemer2006evidence} $a(B) = (1722 + 1.52B/G)\left(1 - \frac{\Delta B}{B/G - B_0}\right)$ for the scattering length conversion, where $\Delta B = 28.72$ and $B_0=-11.60$ is adjusted to shift the zero-crossing to the measured value. The interaction strength is determined as $g =\sqrt{8\pi}a/l_z$, where $l_z = 208(1)~$nm is the vertical harmonic oscillator length. The uncertainty ($\pm \delta g$) in $g$ is primarily contributed by the uncertainty in the magnetic field at the zero-crossing of scattering length. Within the range of reported interaction strengths $-0.004 \geq g_f \geq -0.034$, we have $0.0005 \lesssim \delta g \lesssim 0.0007$.

\subsubsection{Solitary wave characterization}\label{SM:soliton_char}
\begin{figure}[t]
\centering
\includegraphics[width=0.8\columnwidth]{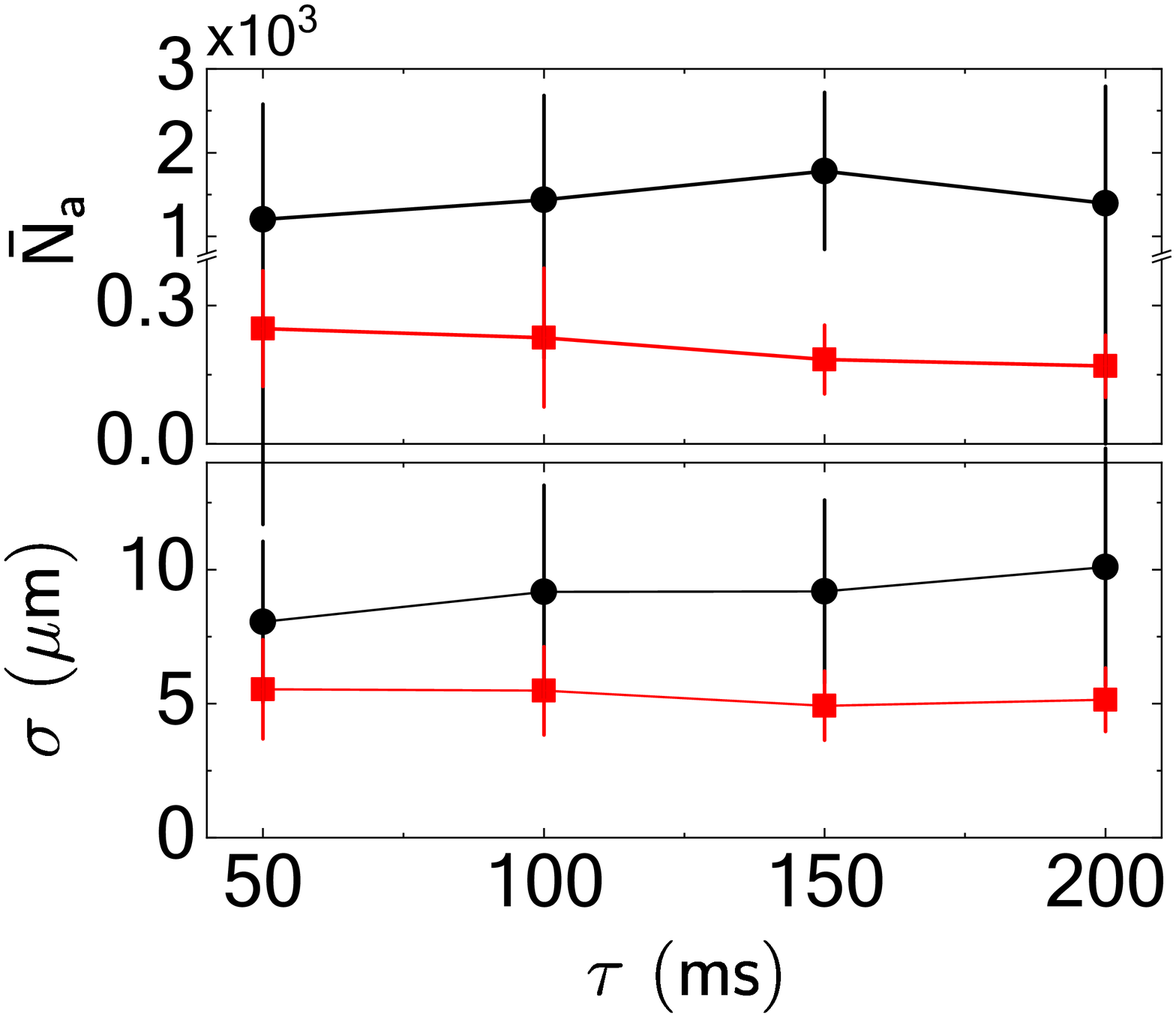}
\caption{Examples at long hold time. Mean atom number $\bar{N}_a$ and root-mean-square diameter $\sigma$ versus hold time $\tau$ measured at $g_f= -0.004$ (black circles) and $-0.019$ (red squares), respectively. Error bars are standard errors.}
\label{figSM:soliton}
\end{figure}

To identify and characterize a solitary wave at long hold time in samples as shown in Fig.~\ref{fig:demo}, we implement 2D peak searching algorithms to search for isolated blobs with peak density above a threshold around $80$\% of the initial density $n_i$, and then fit each blob with a 2D Gaussian in approximation to the Townes profile \cite{chiao1964self}. From each fit, we obtain the root-mean-square (r.m.s.) diameters and extract the atom number $N_a$ under the fitted density profile. We discard fit results that fail to converge in size or have exaggerated aspect ratios that often result from two or more blobs being too close to each other. We then evaluate the mean size $\sigma$ and atom number $\bar{N}_a$. Examples at long hold time $\tau\geq 50~$ms can be found in Fig.~\ref{figSM:soliton}, which shows no signs of rapid collapse or expansion in the mean size and atom number following a single interaction quench.

\subsubsection{Formation of a Townes soliton array}\label{SM:soliton array}
\begin{figure}[t]
\centering
\includegraphics[width=1\columnwidth]{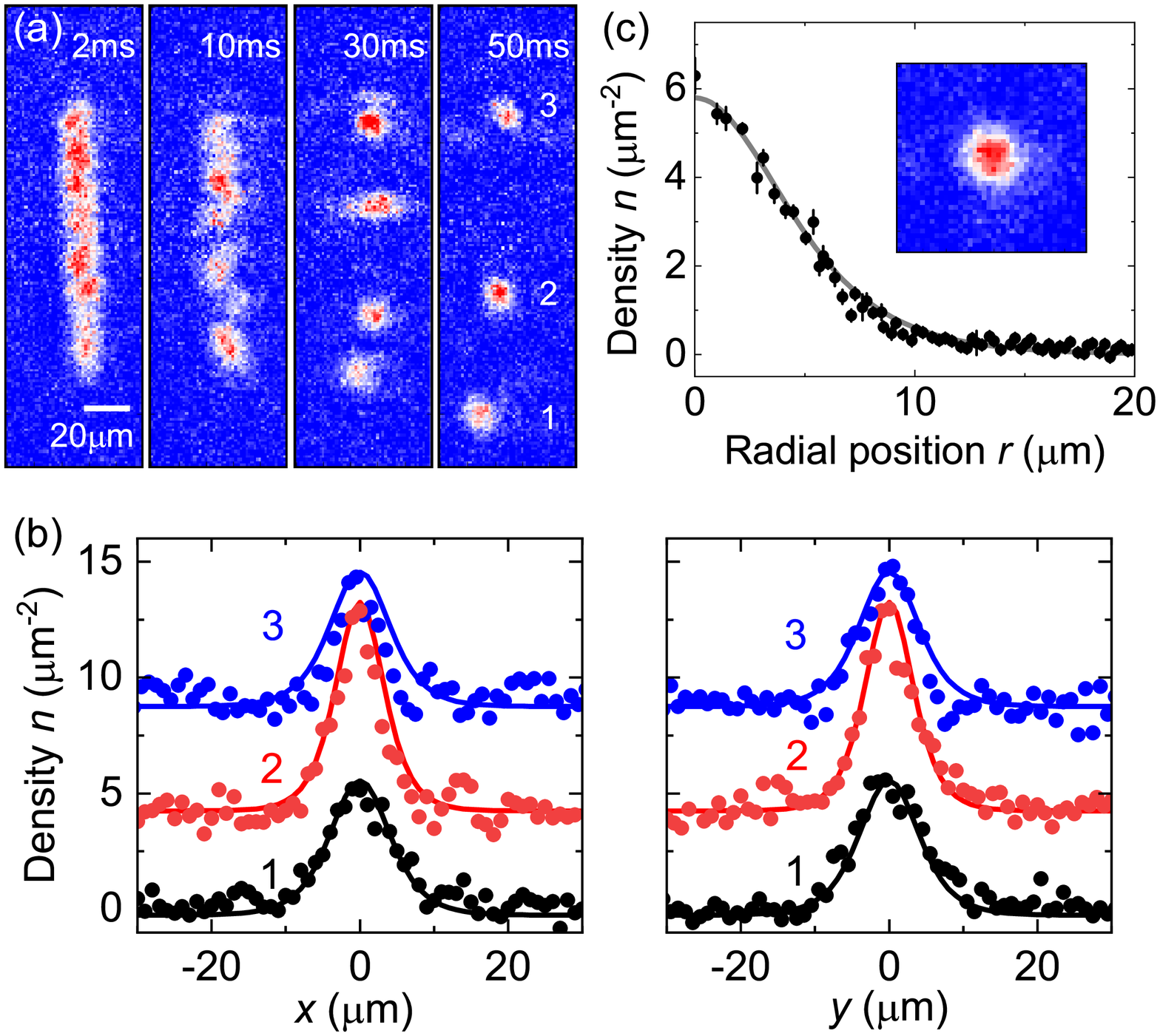}
\caption{Formation of a Townes soliton array. (a) Single-shot images of elongated samples quenched to $g_f=-0.0075$ and held for the labeled time $\tau$. An array of fully isolated solitary waves become visible at $\tau \geq 30~$ms. (b) Density line cuts (solid circles) through the center of three solitary waves as numerically labeled in (a), each offset by $4.5/\mu$m$^2$ for viewing. Solid lines are the Townes profiles of $n_0=5.8/\mu$m$^2$ (for \#1,\#3) and 9$/\mu$m$^2$ (for \#2), respectively. (c) Mean density image of four randomly chosen solitons with $n_0 \approx 5.8/\mu$m$^2$ (inset: $40\times 40~\mu$m$^2$) and the radial profile (solid circles), showing excellent agreement with theory (solid curve).}
\label{fig:demo2}
\end{figure}

To fully demonstrate that Townes solitons can generally form from our quench recipe, we induce MI in another set of 2D samples initially confined and then released from a narrow rectangular wall potential as shown in Fig.~\ref{fig:demo2}(a). We adjust the short side of the samples to be comparable to the MI length scale, so that a \textit{single array} of solitons can form following the interaction quench. This avoids close proximity with many neighboring solitary waves or dispersing blobs, visible in large samples shown in Fig.~\ref{fig:demo}(b). From these narrow samples, we clearly observe well-isolated solitary waves in almost every experiment repetition as shown in Fig.~\ref{fig:demo2}(a) at hold time $\tau \geq 30~$ms. In these solitary waves, we find ubiquitous agreement with the Townes profiles [Fig.~\ref{fig:demo2}(b-c)]. Together with Fig.~\ref{fig:profile}, our observation confirms that Townes solitons can prevail from MI.

\subsection{Quench-induced dynamics in the density power spectrum} \label{SM:sk}
In a 2D gas with uniform mean density distribution, the density power spectrum at finite $k$ is essentially the density static structure factor, which is the Fourier transform of the density-density correlation function \cite{hung2011extracting}. In the following, we discuss the quench evolution of the static structure factor (density power spectrum) measured in our samples, while neglecting density perturbations due to boundary effects.

\subsubsection{Theory of density structure factor after an interaction quench to $g_f<0$}
For a Bose superfluid with initial density $n_i$ immediately following the interaction quench, we expect density waves with wavenumber $0<k<\sqrt{4|g_f|n_i}$ to growth unstably since the usual Bogoliubov dispersion becomes purely imaginary. There is no straightforward theory for evaluating quench evolution at all hold time $\tau$. To gain insights, here we analytically evaluate the quench dynamics only in the very early stage when most of the atoms still remain in the zero momentum state. We focus on the time-evolution of the static structure factor \cite{Hung2013,feng2018coherent}. Analytically, it can be evaluated as
\begin{equation}
S(\mathbf{k})= \frac{1}{N} \sum_{\mathbf{q},\mathbf{q}'} \langle \hat{a}^\dagger_{\mathbf{q}+\mathbf{k}} \hat{a}_{\mathbf{q}} \hat{a}^\dagger_{\mathbf{q}'-\mathbf{k}} \hat{a}_{\mathbf{q}'}\rangle, \label{sa}
\end{equation}
where $\hat{a}_\mathbf{k}$($\hat{a}^\dagger_\mathbf{k}$) stands for the annihilation (creation) operator of a momentum state $|\mathbf{k}\rangle$ and $N$ is the total atom number. At very short hold time $\tau \ll \gamma^{-1}$, the Bose gas is still primarily populated by ground state atoms ($\hat{a}^{(\dag)}_{0}\approx \sqrt{N}$), where $\hbar\gamma = \hbar^2 n_i |g_f|/m$ is the interaction energy, $m$ is the atomic mass, and $\hbar$ is the reduced Planck constant. The structure factor reduces to
\begin{equation}
S(\mathbf{k})=  \langle \hat{a}^\dagger_\mathbf{k} \hat{a}_\mathbf{k} \rangle + \langle \hat{a}_{-\mathbf{k}} \hat{a}^\dagger_{-\mathbf{k}} \rangle + \langle \hat{a}^\dagger_\mathbf{k} \hat{a}^\dagger_{-\mathbf{k}} \rangle + \langle \hat{a}_{-\mathbf{k}} \hat{a}_\mathbf{k} \rangle.\label{sb}
\end{equation}
We perform the following transformation for momentum state within the range $0<|\mathbf{k}|<\sqrt{4|g_f|n_i}$, expressing the momentum state operator with a set of bosonic mode operators $\hat{b}_\mathbf{k}~(\hat{b}^\dagger_{-\mathbf{k}})$ as
\begin{eqnarray}
    \hat{a}_\mathbf{k} &=& i \left[ u_k \hat{b}_\mathbf{k} + v_k \hat{b}^\dagger_{-\mathbf{k}}\right] \nonumber \\
    \hat{a}^\dag_\mathbf{-k} &=& -i \left[ v_k \hat{b}_{\mathbf{k}} + u_k \hat{b}^\dag_\mathbf{-k}\right].
    \label{transform}
\end{eqnarray}
Here, we set the coefficients $u_k = \sqrt{\frac{\hbar\gamma}{2\epsilon(k)}+\frac{1}{2}}$ and $v_k = \sqrt{\frac{\hbar\gamma}{2\epsilon(k)}-\frac{1}{2}}$, $\epsilon(k) = \sqrt{|\epsilon_k^2 - 2\epsilon_k \hbar\gamma|}$ is the imaginary part of the Bogoliubov energy, and $\epsilon_k=\hbar^2k^2/2m$ is the single particle dispersion. $\hat{b}_\mathbf{k}~(\hat{b}^\dagger_{-\mathbf{k}})$ obeys the usual bosonic commutation relation. Using procedures similar to the standard Bogoliubov transformation, we can recast the weakly-interacting Hamiltonian into the following form
\begin{equation}
\hat{H} =\frac{N\mu}{2} + \sum_{\mathbf{k}\neq0} \epsilon(k) (\hat{b}_\mathbf{k}^\dag \hat{b}_{-\mathbf{k}}^\dag + \hat{b}_\mathbf{k} \hat{b}_{-\mathbf{k}} ) - \sum_{\mathbf{k}\neq0} (\frac{\hbar^2k^2}{2m}+\mu ),
\end{equation}
where the summation runs over half of momentum space and $\mu = -\hbar\gamma$ is the chemical potential. Note that, under this transformation, new excitations are generated (and also annihilated) in pairs as hold time increases. In the Heisenbeg picture, these operators obey a set of coupled equations of motion, $\dot{\hat{b}}_\mathbf{k} = \frac{i}{\hbar} [\hat{H},\hat{b}_\mathbf{k}]=- \frac{i}{\hbar}\epsilon(k) \hat{b}^\dagger_{-\mathbf{k}}$ and its Hermitian conjugate, which lead to the following solution
\begin{eqnarray}
\hat{b}_\mathbf{k} &= \hat{b}_{0,\mathbf{k}} \cosh \frac{\epsilon(k)\tau}{\hbar} - i \hat{b}^\dag_{0,-\mathbf{k}} \sinh \frac{\epsilon(k)\tau}{\hbar}\\
\hat{b}^\dag_{-\mathbf{k}} &= \hat{b}^\dag_{0,-\mathbf{k}} \cosh \frac{\epsilon(k)\tau}{\hbar} + i \hat{b}_{0,\mathbf{k}} \sinh \frac{\epsilon(k)\tau}{\hbar},
\end{eqnarray}
and $\hat{b}_{0,\mathbf{k}} (\hat{b}^\dag_{0,-\mathbf{k}})$ is the bosonic mode operator at time $\tau=0$ right after the interaction quench. Plugging this solution into Eq.~(\ref{transform}) and evaluate the structure factor Eq.~(\ref{sb}), we then find the following time-dependent evolution
\begin{eqnarray}
S(\mathbf{k},\tau) &&=
\frac{\epsilon_k}{\epsilon(k)}\left[  (\langle \hat{b}^\dagger_{0,\mathbf{k}} \hat{b}_{0,\mathbf{k}} \rangle + \langle \hat{b}_{0,-\mathbf{k}} \hat{b}^\dagger_{0,-\mathbf{k}} \rangle) \cosh 2 \frac{\epsilon(k)\tau}{\hbar} \right. \nonumber \\
&&- (\langle \hat{b}^\dagger_{0,\mathbf{k}} \hat{b}^\dagger_{0,-\mathbf{k}} \rangle +  \langle \hat{b}_{0,-\mathbf{k}} \hat{b}_{0,\mathbf{k}} \rangle) \nonumber \\
&&\left. -i (\langle \hat{b}^\dagger_{0,\mathbf{k}} \hat{b}^\dagger_{0,-\mathbf{k}} \rangle -  \langle \hat{b}_{0,-\mathbf{k}} \hat{b}_{0,\mathbf{k}} \rangle)\sinh 2 \frac{\epsilon(k)\tau}{\hbar}\right].\label{sc_neg}
\end{eqnarray}
Here, the first line contains mode contributions that are seeded by the initial bosonic mode populations right after the quench. These modes grow `hyperbolically' in the early stage of the quench dynamics. The second and third lines contain the contributions from mode populations that are generated or annihilated from the interaction quench, where the hyperbolic term is expected to vanish, leaving only the constant term (see below).

\subsubsection{Amplification of density waves from density fluctuations prior to the quench} \label{SM:sk_growth}
By using Eq.~(\ref{transform}) and the Bogoliubov transformation, we can further relate the expectation values of the bosonic modes in Eq.~(\ref{sc_neg}) to those of the phonon modes before the interaction quench. We find
\begin{eqnarray}
\langle \hat{b}^\dagger_{0,\mathbf{k}} \hat{b}_{0,\mathbf{k}} \rangle + \langle \hat{b}_{0,-\mathbf{k}} \hat{b}^\dagger_{0,-\mathbf{k}} \rangle &=&\epsilon_k\frac{\hbar(\gamma_i+\gamma)}{\epsilon_i(k)\epsilon(k)} \zeta \label{quenchproj1_neg}\\
\langle \hat{b}^\dagger_{0,\mathbf{k}} \hat{b}^\dagger_{0,-\mathbf{k}} \rangle=\langle \hat{b}_{0,\mathbf{k}} \hat{b}_{0,-\mathbf{k}} \rangle &=&  \frac{\epsilon_k^2+\epsilon_k \hbar(\gamma_i-\gamma)}{2\epsilon_i(k)\epsilon(k)} \zeta,\label{quenchproj2_neg}
\end{eqnarray}
and
\begin{equation}
\zeta = \langle \hat{c}^\dagger_{0,\mathbf{k}} \hat{c}_{0,\mathbf{k}} \rangle + \langle \hat{c}_{0,-\mathbf{k}} \hat{c}^\dagger_{0,-\mathbf{k}} \rangle.
\end{equation}
Here, $\hat{c}_{0,\mathbf{k}}(\hat{c}^\dag_{0,\mathbf{k}})$ is the phonon annihilation (creation) operator, $\epsilon_i(k)= \sqrt{\epsilon_k^2 + 2\epsilon_k\hbar\gamma_i}$ is the Bogoliubov dispersion at interaction $g_i>0$ prior to the quench, $\gamma_i = \hbar n_i g_i/m$, and we have used $\langle \hat{c}^\dagger_{0,\mathbf{k}} \hat{c}^\dag_{0,-\mathbf{k}} \rangle= \langle \hat{c}_{0,\mathbf{k}} \hat{c}_{0,-\mathbf{k}} \rangle=0$ in the above relation since there is no source or sink for phonons in our 2D gas. It is clear that the bosonic mode population Eqs.~(\ref{quenchproj1_neg}-\ref{quenchproj2_neg}) is seeded by the initial thermal phonon population and zero-point fluctuations
\begin{equation}
\langle \hat{c}^\dagger_{0,\mathbf{k}} \hat{c}_{0,\mathbf{k}} \rangle + \langle \hat{c}_{0,-\mathbf{k}} \hat{c}^\dagger_{0,-\mathbf{k}} \rangle = \frac{2}{e^{\epsilon_i(k)/k_B T}-1}+1=\coth \frac{\epsilon_i(k)}{2k_B T}.
\end{equation}
Using the above relations and keeping only the wavenumber $k$-dependence, Eq.~(\ref{sc_neg}) can now be simplified as
\begin{equation}
S(k,\tau) = S_0(k) \left[ 1 +  \frac{2\epsilon_k \hbar(\gamma_i + \gamma)}{\epsilon(k)^2} \sinh^2\frac{\epsilon(k)\tau}{\hbar}\right],\label{sktau}
\end{equation}
where the overall factor
\begin{equation}
    S_0(k)=\frac{\hbar^2 k^2}{2m \epsilon_i(k)} \coth \frac{\epsilon_i(k)}{2k_B T}
\end{equation}
is exactly the equilibrium static structure factor \cite{pitaevskii2016bose} right before the interaction quench.

\subsubsection{The scaling behavior} \label{SM:sk_exp}
In Eq.~(\ref{sktau}), the second term in the bracket represents contributions from the MI-amplified density waves, suggesting density waves at wavenumber $k_p=\sqrt{2|g_f|n_i}$ have the largest amplification rate $\epsilon(k_p)/\hbar=\gamma$. Thus, at short hold time $\tau \ll \gamma^{-1}$, we expect the growth of density power spectrum $\tilde{S}(k_p,\tau) \equiv S(k_p,\tau)/S(k_p,0)$ to obey the following scaling relation
\begin{equation}
    \mathcal{S}(\tilde{\tau}) = \zeta \left[ \tilde{S}(k_p,\tilde{\tau}) -1 \right],\label{EqSM:universal2}
\end{equation}
where $\tilde{\tau} = \gamma \tau$ is the scaled time, $\zeta=\frac{\gamma}{\gamma_i+\gamma}$ is a dimensionless amplitude scaling factor, and $\mathcal{S}(\tilde{\tau})$ is the scaled spectrum; $\zeta \approx \gamma/\gamma_i$ when $\gamma_i \gg \gamma$. The scaled spectrum should display a universal hyperbolic growth at short hold time
\begin{equation}
    \mathcal{S}(\tilde{\tau}) = 2 \sinh^2(\tilde{\tau}).
\end{equation}

\subsubsection{Relationship with quench dynamics at $g_f>0$}
We note that the time-dependent density power spectrum Eq.~(\ref{sktau}) is essentially the analytical continuation of quench-induced Sakharov oscillations in the static structure factor at $g_f>0$ \cite{Hung2013},
\begin{equation}
    S(k,\tau) = S_0(k) \Big[ 1 + \frac{\epsilon_i(k)^2-\epsilon^2(k)}{\epsilon(k)^2} \sin^2 \frac{\epsilon(k)\tau}{\hbar} \Big],
\end{equation}
where now a coherent, sinusoidal oscillation in the structure factor replaces the hyperbolic growth. The calculation is in principle valid for all hold time $\tau$ for $g_f>0$, in contrast to the case of $g_f<0$, provided no global trap effect or other effects set in.

\subsubsection{Experimental test of the scaling behavior in the density power spectrum}
We experimentally test the scaling behavior Eq.~(\ref{EqSM:universal2}) over an extended time period. We have empirically searched for the best amplitude scaling relation $\zeta$ for a larger time range $0<\tilde{\tau}<5$. We find that the spectra scale the best with $\zeta \propto \gamma$, keeping explicit dependence on $n_i|g_f|$. We adopt a simple amplitude scaling factor $\zeta = \gamma/\bar{\gamma}_i$, using mean $\bar{\gamma}_i=\hbar \bar{n}_i g_i/m=306$~s$^{-1}$ evaluated from all samples (mean $\bar{n}_i=6/\mu$m$^{2}$ and standard deviation $\delta n_i = 1/\mu$m$^{2}$). We note that an alternative amplitude scaling factor $\zeta = \gamma/(\bar{\gamma}_i+\gamma)$ close to the exact form in Eq.~(\ref{EqSM:universal2}) gives a similar data collapse.

\subsection{Soliton collision dynamics}\label{SM:loss}
\subsubsection{Soliton binary collision coefficient $\Gamma_s$}
Binary collisions between solitons can lead to merger \cite{nguyen2014collisions}, which makes soliton atom number $N_a > N_\mathrm{th}$ much greater than the Townes threshold and induces collapse and rapid atom loss. Soliton binary loss behavior can be effectively described by $\dot{N}_s/N_s =-\Gamma_s n_s$, where $N_s$ is the number of solitons, $n_s=N_s/A$ is the surface density, and $\Gamma_s$ is the 2D binary loss coefficient as discussed in the main text. For total atom number loss dominated by soliton binary collision loss, we should have $\dot{N}/N =-\Gamma_\mathrm{2body} n$, where $n$ is the atom number density. We convert the soliton number into total atom number $N$ using $N_s = N/\bar{N}_a$ \cite{nguyen2017formation,everitt2017observation} and assuming $\bar{N}_a$ is approximately constant over the experiment time. We find the simple relation $\Gamma_s = \Gamma_\mathrm{2body} \bar{N}_a$, relating the measured loss coefficient $\Gamma_\mathrm{2body}$ to the binary collision coefficient $\Gamma_s$.

\subsubsection{Collision lifetime of solitons}
Following the determination of universal collision dynamics, we also obtain the collision lifetime of solitons right after the formation process near $\tilde{\tau}\approx \tilde{\tau}_c$. We calculate the lifetime $\Delta \tau_s = \frac{1}{\Gamma_s n_s} \approx \frac{m \bar{N}_a}{\eta \pi \hbar n_i} \approx \gamma^{-1}$, where we have used $n_s = n_i/\bar{N}_a$, the measured universal threshold $\bar{N}_a|g_f| = 6$ (Fig.~\ref{fig:demo}), and the measured constant $\eta= 1.5$ (Fig.~\ref{fig:loss}).

\subsection{Collapse and expansion dynamics of an unstable Townes soliton}
This section is dedicated for gaining further information about the evolution of unstable Townes solitons. We determine their collapse and expansion time scales by driving the atom number more than three times away from the Townes threshold via a second interaction quench.

\begin{figure}[t]
\centering
\includegraphics[width=0.8\columnwidth]{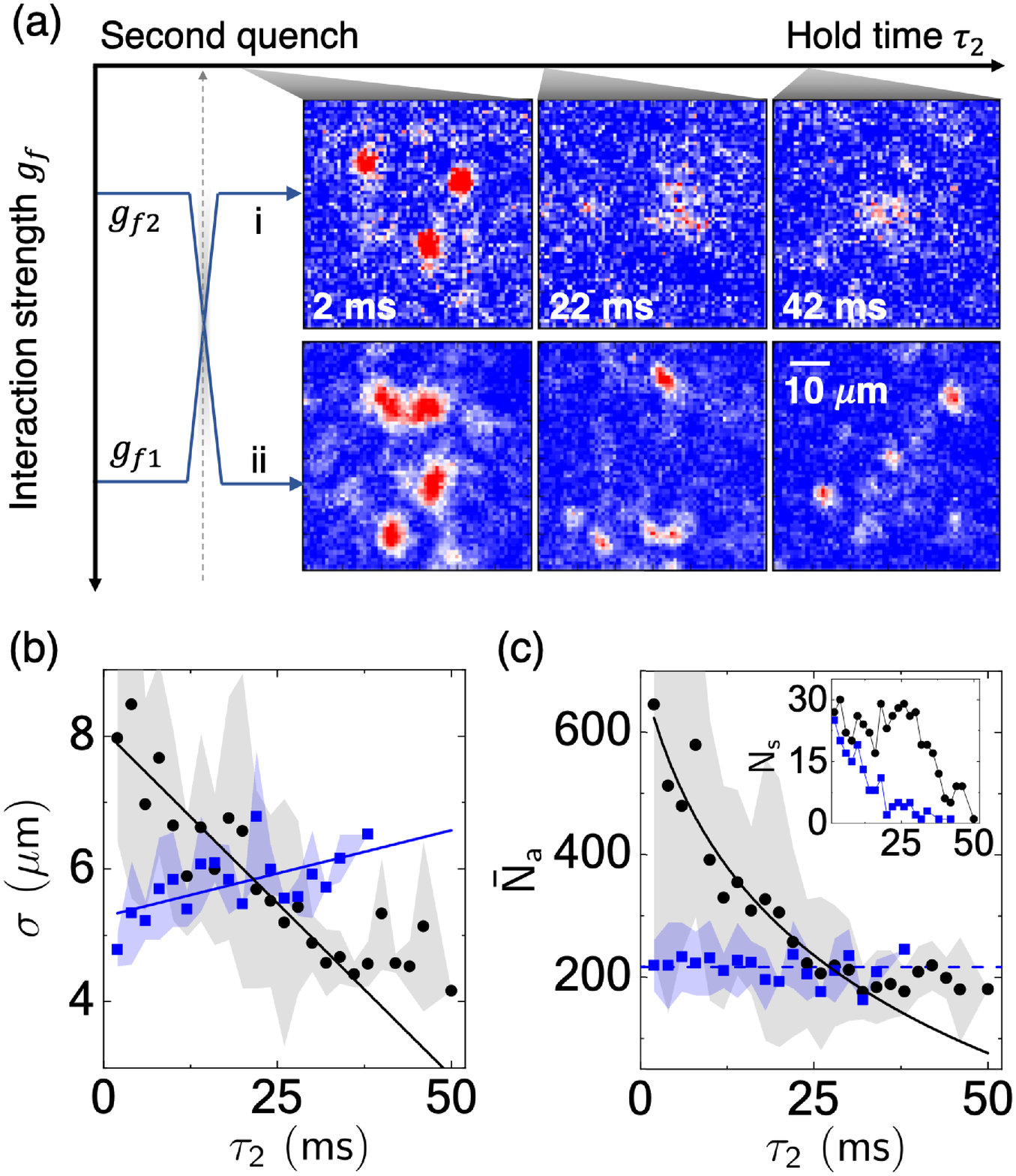}
\caption{Collapse and expansion dynamics following second interaction quenches. (a) Path i (ii): expansion (collapse) dynamics is initiated in solitons that are initially formed at $g_{f1} = -0.027$ (or $g_{f2} = -0.008$) followed by a second quench to $g_{f2}$ (or $g_{f1}$). Single-shot sample images are recorded at the indicated hold time $\tau_2$ after the second quench. (b-c) mean size $\sigma$ and atom number $\bar{N}_a$ versus $\tau_2$ for paths i (blue squares) and ii (black circles), respectively. In (b), solid lines are linear fits, giving a collapse (expansion) rate $\dot{\sigma} /\gamma' \approx -1.8~\mu$m ($0.5~\mu$m) normalized by the interaction energy unit $\gamma' = \hbar \bar{n} |g_{f2}-g_{f1}|/m \approx 55~$s$^{-1}$ and $\bar{n} \approx 6/\mu$m$^{2}$ is mean initial peak density. In (c), blue dashed line marks the Townes threshold $N_\mathrm{th}=5.85/|g_{f1}|$. Black solid line is a guide to the eye, given by a tentative three-body loss fit (see text). Inset shows the observed soliton number $N_s$ from the ensemble measurements. Shaded bands represent standard errors.}
\label{figSM:2ndquench}
\end{figure}

\subsubsection{Second interaction quench}
We apply a second interaction quench after solitons form at a sufficiently long hold time $\tau=50~$ms, hold for an additional time $\tau_2$, and perform imaging. In a quench path labeled (i) in Fig.~\ref{figSM:2ndquench}(a) to a less attractive interaction, we induce immediate soliton expansion, during which $\bar{N}_a$ remains constant, as expected, but the number of solitons $N_s$ observed from the ensemble measurements greatly reduces; see Fig.~\ref{figSM:2ndquench}(c) and inset. For a reversed quench path (ii) to a more attractive interaction, solitons collapse. In a short time scale $\gamma'^{-1} \sim 20~$ms corresponding to the interaction energy difference between the two quenches, we observe rapid reductions in both $\sigma$ and $\bar{N}_a$.

In the collapse dynamics, we tentatively attribute the atom number loss within a soliton to few-body inelastic collisions that are primarily due to three-body recombination. However, in Fig.~\ref{figSM:2ndquench}(c), we obtain an unphysical loss coefficient that is five orders of magnitude larger than that measured in thermal samples \cite{kraemer2006evidence}. A rapid three-body loss rate was also reported in the collapse within 1D solitons in a related experiment setting~\cite{nguyen2017formation}. We believe that a more likely explanation for the rapid loss may be due to a combination of three-body loss under higher local collapse density and collective matter-wave ejection out of the solitons during the collapse \cite{Kagan1998Collapse,donley2001dynamics,Saito2001Collapse,saito2002mean}, which is challenging to detect given limited image resolution and signal-to-noise. Interestingly, the collapse seemingly slows down as $\bar{N}_a$ drops to the new threshold value. We suspect either the atom loss has regulated the collapse, perhaps due to collective wave emission, or solitons with initial $N_a$ close to the new threshold survive. From the inset of Fig.~\ref{figSM:2ndquench}(c), the observed continuing decrease of soliton number beyond $\tau_2 \geq 36~$ms may hint more of the latter case.

In either quench paths (i) and (ii), soliton evolution is clearly visible within the interaction time scale $\gamma'^{-1}$. For the dynamics after just a single quench, such rapid evolutions in mean atom number and soliton size are not observed. We therefore conclude that those surviving solitons formed by MI in a single quench are quasi-stationary within our experiment time $<200~$ms as their norm are sufficiently close to the Townes threshold.

\subsubsection{Fitting atom number loss during rapid soliton collapse}\label{SM:3body}
Following the second interaction quench in path (ii) as shown in Fig.~\ref{figSM:2ndquench}, mean atom number in a soliton $\bar{N}_a(\tau_2=0)$ exceeds the new Townes threshold by three-fold and solitons begin to collapse. From the linear fit in Fig.~\ref{figSM:2ndquench}(b), we obtain an approximate linear time-dependent size $\sigma(\tau_2) = \sigma_0 + \dot{\sigma}\tau_2$, where $\sigma_0=8.0~\mu$m and $\dot{\sigma}=-0.11~$mm/s. Here we consider a simple case where the atom number loss within a soliton is fully due to three-body recombination. Because the vertical oscillator length $l_z=208~$nm far exceeds the magnitude of the 3D scattering length $|a|=1.1~$nm, we expect the three-body loss behavior to be 3D in nature. To quantify the three-body loss rate, we develop a model that takes into account the shrinking soliton size in the 2D plane, while assuming that the vertical wave packet remains in the harmonic ground state. This is justified because the 2D density $n$ needs to increase by $\sim170$ times for the interaction energy to approach the vertical trap vibrational energy. Such dramatic increase in 2D density is not observed in our images, where the soliton size remains larger than the image resolution during the time of collapse. We therefore adopt the standard three-body recombination loss model
\begin{eqnarray}
    \frac{d \bar{N}_a}{d\tau_2} = -\frac{L_3}{9\sqrt{3}\pi^3\sigma(\tau_2)^4l_z^2} \bar{N}_a^3,\label{EqSM:3body}
\end{eqnarray}
where $L_3$ is the three-body loss coefficient and we have used a Gaussian form to approximate the soliton 3D density profile. From the above equations and the approximate linear time-dependence in $\sigma$, we derive an analytical formula
\begin{equation}
    \bar{N}_a(\tau_2) =\frac{\bar{N}_a(0)}{\sqrt{1+ \frac{2L_3\bar{N}_a(0)^2}{27\sqrt{3}\pi^3l_z^2 |\dot{\sigma}|}\left[\frac{1}{(\sigma_0 +\dot{\sigma}\tau_2)^3} - \frac{1}{\sigma_0^3} \right]}},\label{EqSM:3bodycollapse}
\end{equation}
where $L_3$ is the fit parameter. This effective three-body loss model appears to fit our data except beyond $\tau_2>36~$ms when the atom number reaches the new Townes threshold. The fitted $L_3=1.1(1)\times 10^{-23}~$cm$^6/$s is nonetheless five orders of magnitude larger than that measured in a thermal sample \cite{kraemer2006evidence}, which we believe is not physical. To reconcile this discrepancy may require a different collapse dynamics that, for example, creates three-body loss under higher local collapse density beyond our image resolution and also rapidly ejects atoms out of a collapsing soliton \cite{Kagan1998Collapse,donley2001dynamics,Saito2001Collapse,saito2002mean}.

\end{document}